Research Article

# An Analytical Approach for Preliminary Response-Based Design of Semi Rigid Moment Resisting Frames

Arash Shafegati Charvade[1*], Seyyed Mostafa Mousavi Janbeh Sarayi [2*]

[1] Department of Civil, Chemical, Environmental, and Materials Engineering, University of Bologna (UNIBO), Italy
[2] Department of Mechanical Engineering, University at Buffalo, USA

**Abstract**

In many cases, beam to column connections in structural frames are semi rigid, but they are considered to be ideally rigid or pinned due to their computational complexity and shortage of designing methods. In this paper, connections are considered as linear rotational springs with variable stiffness. Moreover, the main contribution of the present study is a preliminary design method of moment resisting frames, considering semi rigid behavior of connections, by means of presented diagrams for different frames. These diagrams relate three features of a frame together; stiffness of frame's connections, geometrical properties of frame's elements, and its lateral displacement. These diagrams can be used when the required stiffness of frames connections is needed while the desired response of the frame, dimensions of the frame and the ratio of second moment of inertia of its elements are known. On the other hand, they could be used to obtain the ratio of beams length to columns length and the ratio of second moment of inertia of beams to columns alongside the stiffness of frame's connections while the only known data is the number of frame's grids in X and Y directions and its desired response.

## 1. Introduction

Connections are a principal part of structural frames as they have significant effects on frame's response. In conventional analysis of structural frames, connections are considered to be ideally pinned or fully rigid in favor of simpler calculations. However, the fact that almost all types of connections have rotational stiffness [1] have them directly influence the analysis procedure and the results. Therefore, designers are to take the stiffness of connections into account in their calculations as ignoring it gives unreliable results.

In AISC 360-16 [2], two major types of connections are defined; simple and moment resisting connections. Moment resisting connections are categorized into fully restrained (FR) and partially restrained (PR) connections. This classification relies on the region of the moment rotation curve that connections behavior fits into. Employing partially restrained connections, also known as semi rigid connections [3], changes the moment distribution along beams and columns and, due to second-order effects, increases frame's drift [4,5].

A number of researches [6]–[10] have conducted experimental tests on connections to investigate their moment rotation response. Although these tests and their results come in handy when it comes to designing connections, all types of connections could not be covered, all obtained moment rotation diagrams are not reliable due to test conditions [11], conducting tests and obtaining the moment rotation relationship for all connections in full scale is expensive [12]. Analytical methods have always been of interest in analyzing structural elements and frames [13]–[16]. Furthermore, semi rigid connections in moment resisting frames have been studied in details as a single part and the results are represented as moment rotation diagram [17]–[21]. The downside of these results is that they are incapable of predicting the response of entire frame due to variation of connections properties. This being the case, Jaspart and Maquoi [22] investigated braced frames with semi rigid connections and described the mode of application of the elastic and plastic response of frames. In another study, Braham and Jaspart [23], using computer simulations of the behavior of a real structure, showed that, limited to the cases where the joints show a high ductility, it is safe to design a building under the assumption of pinned joints even when they show semi rigid behavior. Shi et al. [24] investigated the behavior of beam to column joint rotation in a steel frame and showed that it has significant effect on the internal force distribution and the global deformation response. Recently, Kaveh et al. [25] considered connection types (simple or rigid) and section of elements as design variable for seismic design optimization of steel moment frames.

In this study, initial stiffness of connections is considered as classification index. Two variables, rigidity characterization ($\alpha_1$) and rigidity index ($\alpha_3$), are introduced which appear in parametrical stiffness matrix of frames. These coefficients are used as designing factors in the results. Following this, the displacement vector of frame is calculated parametrically. For generalizing results, normalized displacements are obtained by the maximum displacement of the same frame in case of ideally pinned connections and the same load distribution. This, also, results in omitting loads effect in the results. By increasing the stiffness of connections from zero to very large values and plotting the result, one unique diagram (regardless of applied loads) would be obtained. Each diagram relates three features of a frame; stiffness of the frame's connections, geometrical properties of the frame, and its lateral displacement. They are used if two of these features are defined and the other one is required.

## 2. Materials and Formulations

In order to control the behavior of a frame with semi rigid connections two terms are necessary; stiffness matrix of semi rigid frame and a control pattern to monitor the response of semi rigid frames.

### 2.1. Stiffness Matrix of a Semi Rigid Frame

In order to derive the global stiffness matrix, a general planner frame is considered as shown in Figure 1. All connections of this frame are assumed as linear rotational springs. General methods to derive stiffness matrix of a frame with semi rigid connections are available [22]. To implement these methods on a frame with different types of connections and different stiffness, a frame should be divided into appropriate number of sub frames that each of them has the same connections.

Generally, stiffness of a semi rigid frame is related to the modulus of elasticity (E), moment of inertia (I), area of cross sections (A) lengths of beams and columns (L) and the value of stiffness of its connections ($K_s$):

$$K \sim f(E, I, A, L, K_s) \qquad (1)$$

where it is assumed that beams and columns material are the same with different geometrical properties ($E_c = E_b = E$).

Herein, to calculate the stiffness matrix, geometrical properties of columns ($A_c, L_c, I_c$) and geometrical ratios ($\rho_A, \rho_L, \rho_I$) are used:

$$\rho_A = \frac{A_{beam}}{A_{column}} \qquad (2a)$$

$$\rho_L = \frac{L_{beam}}{L_{column}} \qquad (2b)$$

$$\rho_I = \frac{I_{beam}}{I_{column}} \qquad (2c)$$

so Eq. (1) can be re-written as

$$K \sim f(E, A_c, L_c, I_c, \rho_A, \rho_L, \rho_I, K_s) \qquad (3)$$

In the Appendix, the semi rigid stiffness matrix for a one storey one bay semi rigid frame (Figure 2a) is given as an illustration. As it can be observed, there are many entries that stiffness matrix depends on, so it would be difficult and lengthy to represent results for all of them. In the present study, two other coefficients, appearing in the stiffness matrix, are introduced so that the results could be presented practically and extensively based on them. These coefficients are α_1, which is called rigidity characterization, and α_3, which is called rigidity index:

$$\alpha_1 = \frac{K_s}{K_b} \qquad (4)$$

$$\alpha_3 = \frac{\rho_L}{\rho_I} \qquad (5)$$

As $\alpha_1$ is the ratio of the connections stiffness to the beams stiffness and $\alpha_3$ is the ratio of dimensions of frame, using these coefficients to represent the results make them include an extensive geometrical and mechanical properties.

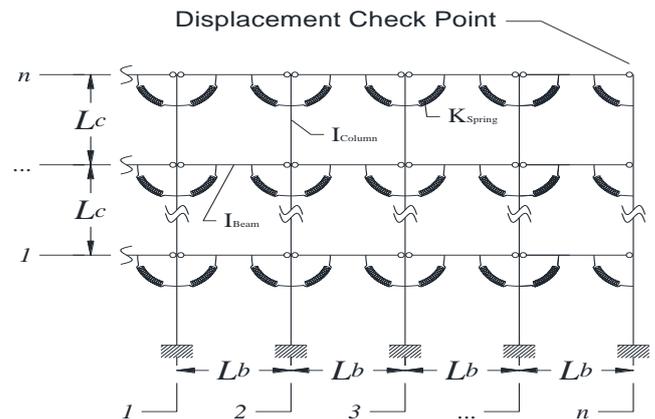

**Figure 1.** Planner frame with connections as rotational springs

### 2.2. Monitoring the Behavior of A Semi Rigid Frame

To monitor the behavior of a semi rigid frame, a displacement control pattern is required. This pattern could be the relative rotation of a vertical cross section of the beam to the face of the column at the connection zone which already there are many formulas and data banks available that have been developed by different researches [23]. In this paper, the maximum lateral displacement at the top right corner of a planner frame (displacement checkpoint in Figure 1) is used as the control pattern of displacement. Herein, Lateral displacement of a frame varies between two boundaries which the lower one is related to the ideally rigid connections and the upper one to the ideally pinned connections. Displacement of a frame between these two boundaries is considered as semi rigid displacement of that frame. Displacements of a frame are normalized by the maximum amount of displacement of the same frame meaning that displacement of a semi rigid frame is divided by the maximum displacement happening at the check point of the same frame when the connections are ideally pinned. Normalized displacement Nv helps to have generalized results and to exclude the variation of lateral or vertical forces in calculations since the load distribution in two cases are the same.

Value of normalized displacement only depends on the ratio of the stiffness of the frame to the stiffness of the same frame with ideally pinned connection; this is illustrated in Eq. (6). Nv is in (0,1] domain and the response of different frames could be compared by the value of Nv in this domain.

$$N_v = \frac{u_{\alpha_1}}{u_{\alpha_1=0}} \sim \frac{(K_{\alpha_1})^{-1} \times (F)}{(K_{\alpha_1=0})^{-1} \times (F)} \quad (6)$$

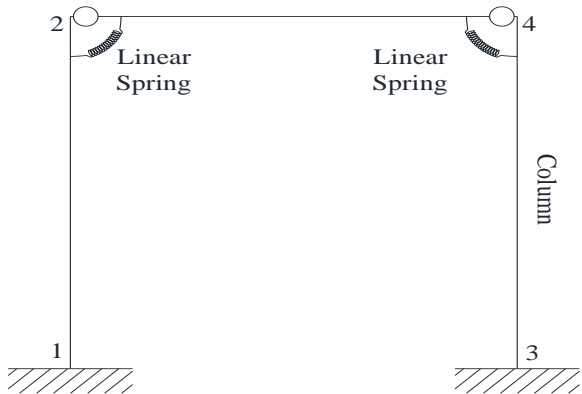

(a)

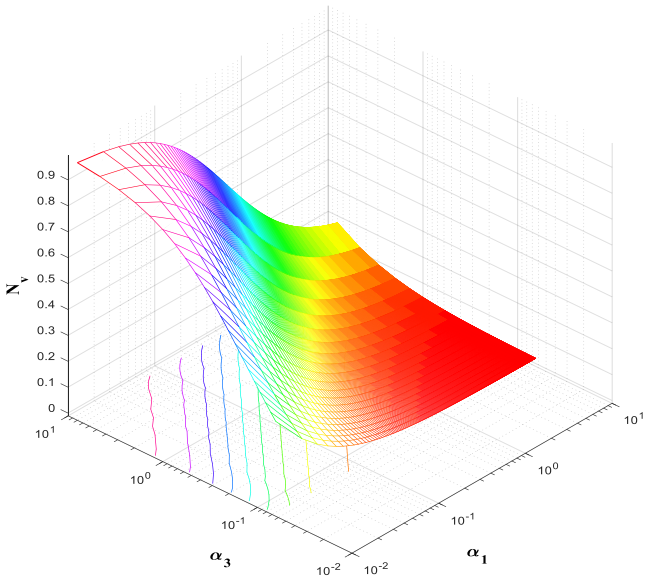

(b)

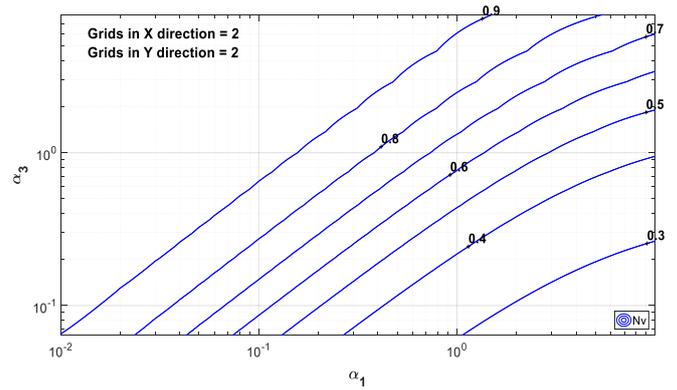

(c)

**Figure 1.** a) 3D surface of Nv with respect to α1 and α3. b) Diagram related to a one storey, one bay simple frame c) 2D diagram related to the planar frame with one storey and one bay

## 3. Numerical Discussion

To calculate displacement of a frame, the well-known formula of Eq. (7) is employed:

$$Displacement = Stiffness^{-1} \times Load \quad (7)$$

As mentioned, in this study the normalized displacement is in non-dimensional form $N_v$ and α_1 and $\alpha_3$ are considered as designing factors. $\alpha_3$ is defined in order to combine dimensions of a frame and its elements (geometrical properties of beams and columns) so that they could be represented as one parameter in the results. In Table.1 the values of $\alpha_3$ for different cases of beams and columns are presented. This table can be extended for a large number of different W-sections as this parameter is an essential part of design. Using this table, one can understand an amount of $\alpha_3$ corresponds to what type of beam and column.

The results of present study, in their actual form, are 3D graphs in which the values of normalized displacements are plotted in form of a 3D surface with respect to $\alpha_1$ and $\alpha_3$ as it can be seen in the Figure 2b which is the related diagram for a one storey, one bay frame that its stiffness matrix is presented in Appendix. $\alpha_1$ and $\alpha_3$ are in logarithmic scale. In order to make the diagrams readable, the 3D surface of $N_v$ is projected on the 2D surface of α_1 and α_3 in 9 discretized lines (10% umax, 20% umax, …,90% umax). Accordingly, for each Nv the relation between α_1 and α_3 is obtained. This helps to have a 2D diagram with α_1 in horizontal direction and $\alpha_3$ in vertical direction with 9 parallel lines that each pair represents a unique value of Nv (Figure 2c).

For more clarification and checking the reliability of results, a model of a 4 storeys, 5 bays frame with semi rigid connections (Figure 3) is studied and the results are compared with the results of FEM method drawn from Etabs program. All required properties are available in Table 2. This frame is studied for different amount of normalized values $N_v$ (0.4, 0.5, 0.6, and 0.7). A brief report of this verification is given in Table 3. It can be seen that the results of present study are in excellent agreement with FEM method results.

The diagrams of present study can be used in two different cases. First case happens when dimensions of a frame and its elements cross sections are known and the stiffness of its connections are required in order to have a desired response. Second case happens when the number of grids in X and Y directions of a frame and the response of the frame are known and dimensions of the frame and elements cross sections and stiffness of its connections are required. Table 4 shows available and required parameters in each case of study and detailed explanations are provided as follows. In supplementary material, thirty diagrams for different frames with different number of grids in X and Y directions are presented. Grids Numbering in X and Y direction is shown in Figure 4. It should be noted that grids numbering in Y direction starts from the ground level, In fact, number of bays and number of storeys are one less than number of grids in X and Y direction respectively.

**Table 1.** Values of $\alpha_3$ for different sections and lengths

| | | Beam | | | | | | | |
|---|---|---|---|---|---|---|---|---|---|
| | | W690X192 | W460X235 | W530X182 | W310X202 | W360X147 | W410X149 | | |
| $I_x$ ($10^6 mm^4$) | | 1980 | 1270 | 1230 | 516 | 462 | 620 | | |
| | L(mm) | 300 | 400 | 500 | 600 | 700 | 800 | | |
| 4450 | 500 | 1.35 | 2.80 | 3.62 | 10.35 | 13.48 | 11.48 | W610X455 | |
| 2840 | 400 | 1.08 | 2.24 | 2.89 | 8.26 | 10.76 | 9.16 | W610X307 | Column |
| 2360 | 300 | 1.19 | 2.48 | 3.20 | 9.15 | 11.92 | 10.15 | W610X262 | |

### 3.1. Case I

To illustrate that how this method can be applied in practice, the model of Figure 3 and Table 2 is considered. The aim of this case of numerical model is to calculate the connections stiffness in order to control the displacement of the frame. In this procedure, the value of $\alpha_3$ is needed which can be calculated by means of Eq. (5) ($\alpha_3 = 6.4$). By locating $\alpha_3$ in the related diagram that is given in Figure 5, the value of $\alpha_1$ for each amount of Nv can be read out. Later on, $\alpha_1$ will be used to calculate the amount of stiffness that is required in the frames connections in order to have the desired Nv. So far, three values of $N_v$ (desired response of the frame), $\alpha_3$ (calculated by means of Eq. (5) according to the frames dimensions) and $\alpha_3$ (obtained from related diagram) are known. In follow, by means of Eq. (4), required stiffness of connections in order to have the considered $Nv$ can be calculated. Results could be found in Table 5. The value of $\alpha_3$ would remain constant as far as the dimension of the frame and the geometrical properties of frame's elements do not change.

**Table 2.** Elements and its properties used in the frame of the example

| Elements | Type | $I$ ($m^4$) | $L$ ($m$) | $\rho_L$ | $\rho_I$ |
|---|---|---|---|---|---|
| Column | HE300E | 5.92e-04 | 3.2 | 2.5 | 0.4 |
| Beam | IPE400 | 2.31e-04 | 8.0 | | |

### 3.2. Case II

In the previous section, dimensions of the frame and the geometrical properties of beams and columns were known while the stiffness of connections were required. It is also possible to evaluate the geometrical properties of a frame and its elements besides the stiffness of its connections ($\rho_I, \rho_L$, Ks) for a desired response of a frame while just number of grids in X and Y directions are known. Since $Nv$ lines relate the values of $\alpha_1$ and $\alpha_3$ together, by moving over each line, a large variation of geometrical properties (related to $\alpha_3$) and stiffness of connections (related to $\alpha_1$) are available. For example, by fixing the value of Nv to 0.4, if the value of $\alpha_3$ is selected in a way that its intersection with $\alpha_1$ happens on the line, a large range of ($\alpha_1,\alpha_3$) couples would be available which return a constant displacement. In Fig. 5, five different values for $\alpha_1$ and $\alpha_3$ is shown with blue dots on the diagram which all are placed on the line related to $Nv= 0.4$. By finding the values of $\alpha_1$ and $\alpha_3$ for each related dots, the ratios of lengths and moments of inertia could be found as it can be seen in Table 6. In this table an optional value for $\rho_L$ is chosen which depends on frames dimensions or the will of the designer. Then, regarding to the values of $\rho_L$, the value of $\rho_I$ can be calculated by means of Eq. (5) according to the value of $\alpha_3$. Consequently, the moment of inertia for beams and columns can be evaluated in a way that their ratio become $\rho_I$. Since the value of $\alpha_1$ is known and the beams moment of inertia is already evaluated by means of $\alpha_3$, the value of Ks could be calculated by means of Eq. (4).

**Table 3.** Comparison between assumed $Nv$ from proposed diagrams and obtained Nv from FE software

| Status of connection | pinned | Semi rigid | Semi rigid | Semi rigid | Semi rigid | rigid |
|---|---|---|---|---|---|---|
| Assumed Nv (from proposed diagrams) | - | 0.7000 | 0.6000 | 0.5000 | 0.4000 | - |
| Calculated stiffness ($Ks$ (N.m/rad)) | - | 893,396.3 | 1,429,434 | 2,203,711 | 3,573,585 | - |
| Displacement at the checkpoint (from FE method for each stiffness) (mm) | 94.9000 | 66.0000 | 56.0000 | 47.000 | 37.000 | 8.000 |
| Obtained $Nv$ (from FE software) | - | 0.6954 | 0.5900 | 0.4931 | 0.3898 | - |

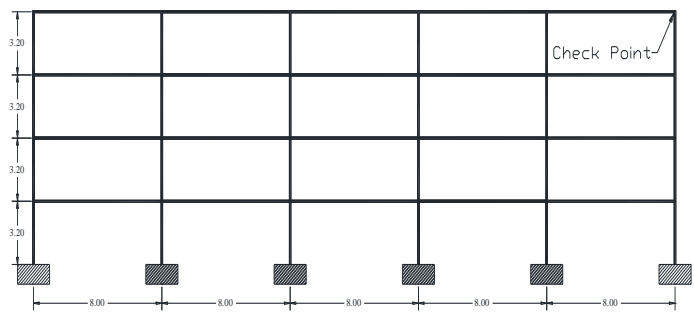

**Figure 3.** Verification model frame, 4 storeys (NY=5) and 5 bays (NX=6) frame, related to the example

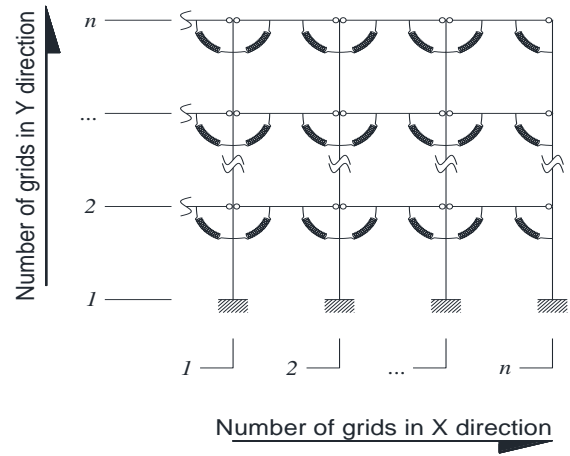

**Figure 4.** Illustration of grids numbering in X and Y direction

**Table 4.** Elements and its properties used in the frame of the

|  | Available Parameters | Required Parameters |
|---|---|---|
| Case I | • Dimension of frame<br>• Dimension of elements cross sections<br>• Response of frame | • Stiffness of connections |
| Case II | • Number of grids in $X$ & Y direction of the frame<br>• Response of frame | • Stiffness of connections<br>• Dimension of elements cross sections<br>• Dimension of frame |

**Table 5.** Calculated $K_s$ for each normalized value

| $Nv$ | $\alpha_1$ $(1/rad)$ | $Ks$ $(N.m/rad)$ |
|---|---|---|
| 0.4 | 0.60 | 3,573,585 |
| 0.5 | 0.37 | 2,203,711 |
| 0.6 | 0.24 | 1,429,434 |
| 0.7 | 0.15 | 893,396.3 |

**Table 6.** Different values of rigidity index and rigidity characterization for one $N_v$.

| $\alpha_1$ $(1/rad)$ | $\alpha_3$ | $\rho_L$ | $\rho_I$ |
|---|---|---|---|
| 0.2 | 2.23 | 2.0 | 0.90 |
| 0.3 | 2.20 | 2.0 | 0.61 |
| 0.4 | 4.40 | 2.0 | 0.45 |
| 0.5 | 5.45 | 2.0 | 0.37 |
| 0.6 | 6.50 | 2.0 | 0.31 |

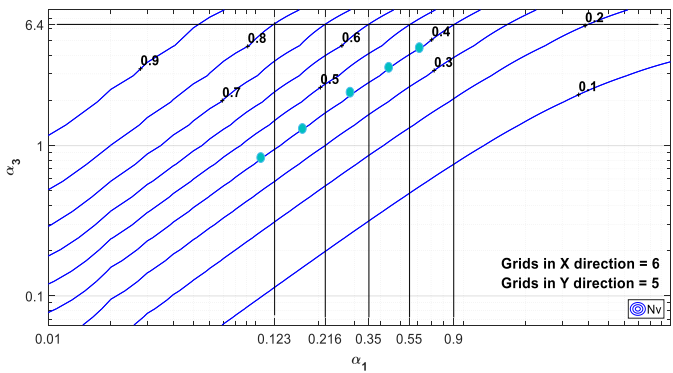

**Figure 5.** Diagram related to a frame with 4 storeys and 5 bays

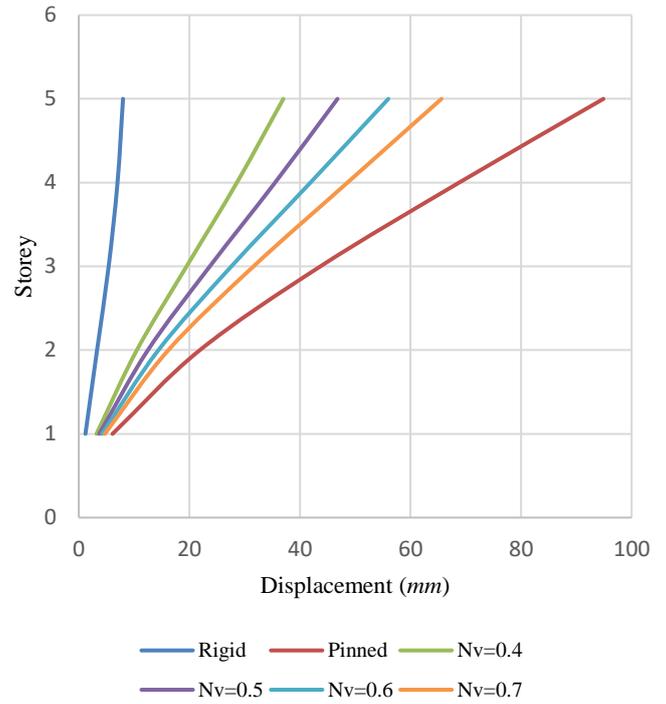

**Figure 6.** Displacements of a frame at checkpoint due to the number storeys and for different $N_v$

## 4. Conclusions

Given that the study of connections is to control the behavior of the entire frame, this research suggests that to obtain stiffness of connections they should be studied as a part of the entire moment resisting frame for a desired response instead of evaluating the stiffness of connections by means of moment rotation method for a single connection. To represent the results of this paper in a simple and practical way, two normalized parameters are used to derive the global stiffness matrix of the frame. These parameters are rigidity index that is the ratio of stiffness of connection to the rotational stiffness of beam and rigidity characterization that is the ratio of the dimensions of frame (length of bay to the height of the storey) to the geometrical properties of frames elements (beams moment of inertia to columns moment of inertia). Obtained diagrams in this research are the illustration of the response of planner moment resisting frames for a large variation of geometrical properties of its beams and columns while the stiffness of its connections tends from zero to a very large value. It is also shown that the response of frame can be precisely controlled with this method since a direct relation between geometrical properties and stiffness of connections is derived. New global model to evaluate the stiffness of connections in a planner frame are proposed in which the normalized value of displacement of a frame with respect to the maximum displacement is suggested. In the supplementary material, related diagrams for thirty different buildings with different number of storeys and bays are provided. Since there is no need for pre analysis or complex calculations in this method and it is just a matter of reading the numerically calculated graphs which covers a large variation of geometrical properties of sections, this method could be a very useful tool for preliminary design of planner frame.

## Appendix

Stiffness matrix of a one storey one bay semi rigid frame:

$$K = E \frac{I_c}{L_c} \begin{bmatrix} k_{1,1} & k_{1,2} & k_{1,3} & k_{1,4} \\ & k_{2,2} & k_{2,3} & k_{2,4} \\ & & k_{3,3} & k_{3,4} \\ Sym & & & k_{4,4} \end{bmatrix}_{12 \times 12} \quad (A1)$$

where we can find the $k_{i,j}$

$$k_{1,1} = \begin{bmatrix} \frac{12}{L_c^2} & 0 & -\frac{6}{L_c} \\ sym & \frac{A_c}{I_c} & 0 \\ sym & sym & 4 \end{bmatrix}_{3 \times 3} \quad (A2)$$

$$k_{1,2} = [0]_{3 \times 3} \quad (A3)$$

$$k_{1,3} = \begin{bmatrix} -\frac{12}{L_c^2} & 0 & -\frac{6}{L_c} \\ 0 & -\frac{A_c}{I_c} & 0 \\ \frac{6}{L_c} & 0 & 2 \end{bmatrix}_{3 \times 3} \quad (A4)$$

$$k_{1,4} = [0]_{3 \times 3} \quad (A5)$$

$$k_{2,2} = \begin{bmatrix} \frac{12}{L_c^2} & 0 & -\frac{6}{L_c} \\ sym & \frac{A_c}{I_c} & 0 \\ sym & sym & 4 \end{bmatrix} \quad (A6)$$

$$k_{2,3} = [0]_{3 \times 3} \quad (A7)$$

$$k_{2,4} = \begin{bmatrix} -\frac{12}{L_c^2} & 0 & -\frac{6}{L_c} \\ 0 & -\frac{A_c}{I_c} & 0 \\ \frac{6}{L_c} & 0 & 2 \end{bmatrix}_{3 \times 3} \quad (A8)$$

$$k_{3,3} = \begin{bmatrix} \frac{12}{L_c^2} + \frac{A_c \rho_A}{I_c \rho_L} & 0 & \frac{6}{L_c} \\ sym & \frac{A_c}{I_c} - \frac{12}{L_c^2} \frac{\alpha_3}{\rho_L^2} \left( \frac{1}{1-4\gamma} - \frac{1}{4\gamma^2-1} + \frac{\gamma}{1-4\gamma^2} \right) & \frac{6}{L_c} \frac{\alpha_3}{\rho_L} \left( \frac{1}{4\gamma^2-1} - \frac{2\gamma}{\gamma-4\gamma^2} \right) \\ sym & sym & 4\left(1 - \frac{3\alpha_3 \gamma}{1-4\gamma^2}\right) \end{bmatrix} \quad (A9)$$

$$k_{3,4} = \begin{bmatrix} -\frac{A_c \rho_A}{I_c \rho_L} & 0 & 0 \\ 0 & \frac{12}{L_c^2} \frac{\alpha_3}{\rho_L^2} \left( \frac{1}{1-4\gamma} - \frac{1}{4\gamma^2-1} + \frac{\gamma}{1-4\gamma^2} \right) & \frac{6}{L_c} \frac{\alpha_3}{\rho_L} \left( \frac{1}{4\gamma^2-1} - \frac{2}{1-4\gamma} \right) \\ 0 & \frac{6}{L_c} \frac{\alpha_3}{\rho_L} \left( \frac{1}{1-4\gamma} - \frac{1}{4\gamma^2-1} + \frac{\gamma}{1-4\gamma^2} \right) & \frac{6\alpha_3}{4\gamma^2-1} \end{bmatrix} \quad (A10)$$

$$k_{4,4} = \begin{bmatrix} \frac{12}{L_c^2} + \frac{A_c \rho_A}{I_c \rho_L} & 0 & \frac{6}{L_c} \\ sym & \frac{A_c}{I_c} - \frac{12}{L_c^2} \frac{\alpha_3}{\rho_L^2} \times \left( \frac{1}{1-4\gamma} - \frac{1}{4\gamma^2-1} + \frac{\gamma}{1-4\gamma^2} \right) & -\frac{6}{L_c} \frac{\alpha_3}{\rho_L} \left( \frac{1}{4\gamma-1} - \frac{2}{1-4\gamma} \right) \\ sym & sym & 4\left(1 - \frac{3\alpha_3}{1-4\gamma}\right) \end{bmatrix} \quad (A11)$$

where

$$\alpha_1 = \frac{K_S}{\alpha_3} \frac{L_c}{EI_c}, \quad \alpha_3 = \frac{\rho_I}{\rho_L}, \quad \gamma = \frac{3}{\alpha_1 \alpha_3} \frac{L_c}{EI_c} + 1 \quad (A12)$$

and the stiffness of beam and column elements are considered as

$$K_{b/c} = \left( \frac{EI}{L} \right)_{beam/column} \quad (A13)$$